\documentclass[a4paper,11pt]{article}
\pdfoutput=1 

\usepackage{jcappub} 

\usepackage[T1]{fontenc} 

\usepackage{amssymb}
\usepackage{amsmath}
\usepackage{subfigure}
\usepackage{pdflscape}
\usepackage{longtable}
\usepackage{enumerate}

\title{Considerations on the radio emission from extended air showers}

\author[a]{E.~Conti,}
\author[b]{G.~Sartori}

\affiliation[a]{INFN, Sezione di Padova, \\Via Marzolo 8, I-35131 Padova, Italy}
\affiliation[b]{Dipartimento di Fisica e Astronomia ``G.Galilei'', Universit\`a di Padova, \\Via Marzolo 8, I-35131 Padova, Italy}

\emailAdd{enrico.conti@pd.infn.it}
\emailAdd{giorgio.sartori@unipd.it}

\abstract{
The process of radio emission from extended air showers produced by high energy cosmic rays has reached a good level of comprehension and prediction. It has a coherent nature, so the emitted power scales quadratically with the energy of the primary particle. Recently, a laboratory measurement has revealed that an incoherent radiation mechanism exists, namely, the bremsstrahlung emission. In this paper we expound why  bremsstrahlung radiation, that should be present in showers produced by ultra high energy cosmic rays, has escaped  detection so far, and why, on the other side, it could be exploited, in the 1--10~GHz frequency range, to detect astronomical $\gamma$-rays. We  propose an experimental scheme to verify such hypothesis, which, if correct, would deeply impact on the observational $\gamma$-ray astronomy.
}


\notoc
\begin{document}
\maketitle
\flushbottom

\section{Introduction}
\label{sec:intro}

The emission of radio frequency (RF) radiation from the Extended Air Showers (EAS) produced by Ultra High Energy Cosmic Rays (UHECR) impinging on the earth atmosphere has been investigated since the 60s, after Askaryan \cite{Ask1962, Ask1965} proposed a coherent mechanism for the production of Cherenkov radiation at MHz frequencies. After the first detection of RF radiation from EAS in 1965 by Jelley et al. \cite{Jelley1965}, many experimental and theoretical efforts have been dedicated to understand its  origin, to characterize it, and to derive informations on the energy and nature of the parent primary cosmic particle. An up-to-date review on these thematics can be found in \cite{Huege2016}. A picture has emerged which describes the radio emission from EAS up to 10~GHz, which can be schematically summarized as follows.

The radiation is produced by two mechanisms. The main one is the so-called geomagnetic effect, due to the separation of positive and negative charges inside the EAS by the earth magnetic field. The second one, which contributes for about 15$\%$ of the emitted power, is the coherent Cherenkov radiation (Askaryan effect), due to the small excess of negative charges in the EAS. For both mechanisms, the emitted electric field is linearly polarized.

The radiation has a coherent nature, hence the energy emitted in RF radiation scales quadratically with the energy $E_p$ of the primary cosmic ray. 
Below $\sim$100 MHz, the coherence condition is satisfied because the RF wavelength $\lambda$ is comparable to the shower thickness. The bunch of particles in the EAS arriving to the detector can be depicted as a dish, with thickness $L_{dish}$  of a few metres and diameter depending of the energy of the primary. 
The requirement for coherence  is  $\nu \lesssim c/L_{dish}\sim$100~MHz, where $c$ is the speed of light.

Above $\sim$100~MHz, another phenomenon ensures coherence under certain conditions.
It has been shown with Monte Carlo simulations \cite{deVries2011, AlvarezMuniz2012, Werner2012} that
the variation of the atmosphere refractive index with the altitude (or better, with air density) provokes the time compression of RF wavefront. This causes the coherence to extend to higher frequency, even above the GHz. The compression, however, occurs only inside a cone with aperture $\psi_c \approx 1^{\circ}$ equal to the Cherenkov angle, and angular spread $\Delta\psi \approx 0.1^{\circ}$. Outside this small angular region, coherence is lost and the RF output power falls dramatically. In this coherence regime, the emitted electric field amplitude $\mathbb{E}$ depends on the frequency and can be modelled, as shown by \cite{ANITA2015,AlvarezMuniz2012}, as
 \begin{equation}
 \label{eq:EvsNu}
\mathbb{E}(\nu)=\mathbb{E}(\nu_0)\exp\left(-\frac{\nu-\nu_0}{\nu_{\tau}}\right)
\end{equation}
with $\nu_0= $ 300~MHz and $\nu_{\tau}\approx $ 500~MHz.

The coherence condition explains why the experimental investigation is mostly concentrated in the frequency range $\lesssim 100$~MHz. 

Although the 10--100~MHz frequency range is the most favourable from the point of view of the emission yield, it is not advantageous with respect to the sky noise. Apart from man-made noise, in that frequency domain the sky is very noisy because of a galactic noise temperature of $10^3$--$10^6$~K \cite{kraus}, which limits the sensitivity of the receiving apparatus. On the opposite, the sky noise is minimal from 1 to 10~GHz, corresponding to the cosmic microwave background radiation temperature of 2.7~K.

The RF radiation is emitted in the forward direction, with small angle with respect to the EAS axis. 
The footprint  at ground is different for the two coherence regimes described before: a circle (for vertical cosmic rays) for $\nu \lesssim 100$~MHz, and an annulus at higher frequencies.

Beside coherent emission, a RF incoherent emission mechanism exists, that is, bremsstralhung radiation (BR), the emission of photons during the deflection of electrons and positrons under the coulomb field of the nuclei. The radiation is neither polarized nor isotropic, having instead a collimated angular pattern. The emission is always present but has been discarded so far as effective mechanism for RF production since the effect scales linearly with $E_p$  and the cross section is low. As we will show in the following, BR emission is indeed negligible for UHECRs but at lower $E_p$ it become comparable or even higher than coherent emission, and could be effective for the detection of astronomical $\gamma$-rays.
Since the BR spectrum is flat over $\nu$, the radiation detection is optimal in the frequency region 1--10~GHz, where the sky noise temperature is minimal.

In this paper, we would like to discuss the BR emission  (section~\ref{sec:BR}) and to compare it with the coherent emission and with the experimental data (section ~\ref{sec:comparison}). In section~\ref{sec:newidea} we examine the possibility to exploit the BR emission in the microwave region for the detection of astronomical $\gamma$-rays and propose a possible experimental scheme. Conclusions are drawn in section~\ref{sec:conclusions}.

Since the emitted RF power $P$ is proportional to the frequency bandwidth $\Delta\nu$ of the receiving apparatus, 
it is more straightforward to deal with physical quantities normalized to $\Delta\nu$. Henceforth, such quantities are marked with an asterisk apex, as, for example, the power per unit frequency $P^* = P/\Delta\nu$, or the electric field per unit frequency $\mathbb{E}^*=\mathbb{E}/\Delta\nu$.

\section{Bremsstrahlung radiation}
\label{sec:BR}
\renewcommand{\thefootnote}{\fnsymbol{footnote}}

The first mention of the BR as an effective mechanism for the generation of RF waves was done in 1966 by Rozental and Filchenkov \cite{rozental}. They showed that the radiation comes mainly from the slow $\delta$-rays produced in the shower, it is isotropic or quasi-isotropic, and estimated that the total radiated energy is $\sim 10^{-14}E_p$ at $\nu \approx$ 25 MHz and for a bandwidth $\Delta\nu/\nu=0.1$. Successively, Jelley and Charman \cite{J1,J2}, pointing out that the radiation is incoherent, argued that its intensity  is negligible compared with that from other (coherent) mechanisms. Thenceforth, BR  was essentially forgotten.

Recently, Conti et al. \cite{noi} measured in laboratory the BR emitted at 11~GHz from bunches of $\sim$100~keV electrons in air. In that experiment, neither Cherenkov nor geomagnetic emission were present. To validate the experimental data, they developed a simple model, where radio frequency is radiated in the  free-electron atomic-nucleus collisions, during the slowdown of the electrons. Effectively, the model  fits  nicely the experimental data. They used the tabulated ``scaled'' cross sections $\sigma_{scaled}$ found in literature, function of the kinetic energy $T$ of the slow electrons, so that the bremsstrahlung RF emission cross section $\sigma(T)$ is:
 \begin{equation*}
\sigma(T) = \frac{Z^2}{\beta^2}\frac{\Delta\nu}{\nu}\sigma_{scaled}
\end{equation*}
where $Z$ is the nucleus atomic number. A Monte Carlo code then followed all particles of the shower generated by the primary electron, and the total  yield was computed. 
According to this model, the emission is not coherent,
 unpolarized, not isotropic but strongly beamed in the forward direction, as a consequence of the angular dependence of the bremsstrahlung cross section.  
At 10~GeV, for instance,  50\% of the radiation is contained within a cone of $2^{\circ}$ semiaperture, but at higher energy the aperture further decreases.
The RF radiation therefore travels along with the EAS charged front, producing, on the receiver detector, a pulse of duration $\Delta T = L_{dish}/c \approx 10$ ns. 

An empirical formula for the emitted power can be obtained by recalling the experimental data and the result of the Monte Carlo simulation of ref. \cite{noi},  and the $\Delta\nu/\nu$ dependence of the cross section:
 \begin{equation}
    \label{eq:BR}
P^*=a~\frac{E_p}{\nu} 
\end{equation}
with $a=2.14\cdot10^{-17}$~ W/MHz if $E_p$ and  $\nu$ are expressed in GeV and GHz, respectively.
Note that the fraction of energy emitted into RF in the bandwidth $\Delta\nu$ is about  $10^{-13}$, not too far from the estimation of ref. \cite{rozental}.

The experimental results in \cite{noi} have been interpreted differently in ref.~\cite{AlSamarai2}: the radio emission is caused by molecular bremsstrahlung coming from the interactions of the very-low energy ionization electrons (kinetic energy $\lesssim 100$ eV) with the molecular species of air. In addition to what happens in EAS, the relatively high plasma density introduces a suppression factor, which arises from the mutual repulsion among the electrons. In such a way, the emitted intensity is reproduced, at least as order of magnitude. To account for the observed angular dependence, the existence of a further term has been speculated,  produced by the oscillations of the electrons in the surrounding electric field they generate. Such radiation has a $\cos^2\theta$ angular shape which would fit well the experimental findings. Although it is difficult to extrapolate that model from laboratory measurements (at high ionization density) to EAS (with low ionization 
density), nevertheless it foresees an isotropic emission for EAS, in total contrast with the model here proposed.
 
The model \cite{noi} is not in conflict with the experimental results of ref.~\cite{Ohta2016} where no RF signal could be detected from a high intensity electron beam (40 MeV energy) shot vertically into air. In fact, the noise of the detection system was much higher than the  BR signal expected from eq.~(\ref{eq:BR}).

\section{Coherent emission}
\label{sec:comparison}

Since the first theoretical works of Askaryan \cite{Ask1962,Ask1965} regarding the coherent Cherenkov emission, and of Kahn and Lerche \cite{Kahn1965} on the interaction of the EAS with the terrestrial magnetic field, many efforts have been devoted to derive an analytical formula which relates the  emitted RF electric field $\mathbb{E}^*$ to the primary particle energy and direction. 
The best attempt was  probably done by  Huege and Falke \cite{HF}, who improved  the previous Allan formula \cite{Allan1970,Allan1971}. 
This approach is however too approximative since it cannot take into accounts all the many features of the radio emission mechanisms. 
 Nowadays scientific community does not usually treat the radio emission with analytical or empirical formulas but relies on computer simulations, where all charged particles of the showers can be followed along their path. Such strategy has been fairly successful in reproducing all recent experimental data  (see for example ref.~\cite{Huege2016}).

\begin{figure}[!]
\centering
\includegraphics[scale=0.95]{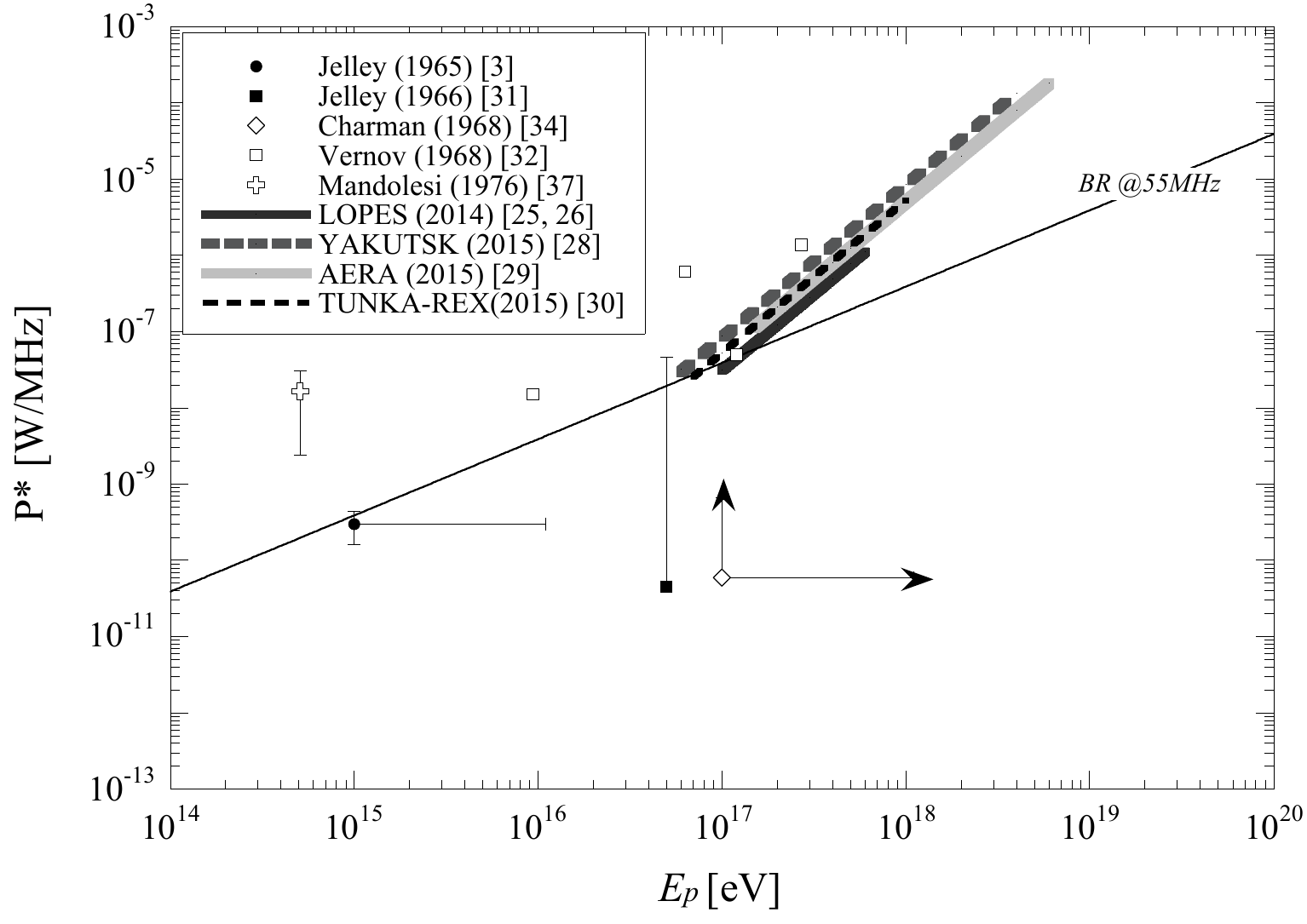}
\caption{Experimental data of the emitted RF power as a function of the cosmic ray energy in the 30--100~MHz range. Also plotted is the BR emission according to eq.~(\ref{eq:BR}), calculated at 55~MHz. 
Arrows indicate that the point is a lower limit. 
}
\label{fg:range1}
\end{figure}

\begin{figure}[t!]
\centering
\includegraphics[scale=0.95]{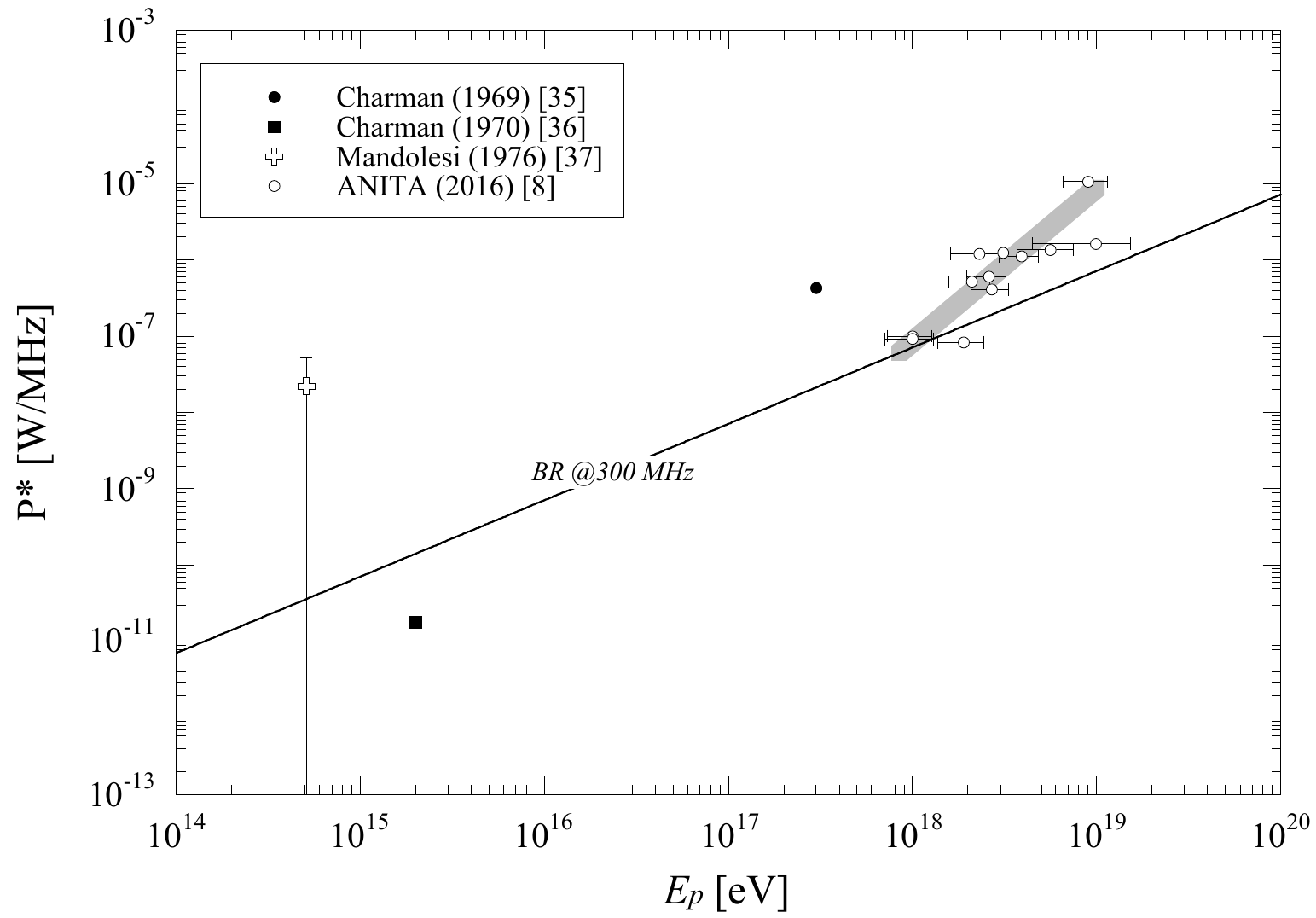}
\caption{Experimental data of the emitted RF power as a function of the cosmic ray energy in the 100--500~MHz range. The  grey band is the fit of the ANITA data according to $P^*\propto E_p^2$. Also plotted is the BR emission according to eq.~(\ref{eq:BR}), calculated at 300~MHz.
}
\label{fg:range2}
\end{figure}

\begin{figure}[t!]
\centering
\includegraphics[scale=0.95]{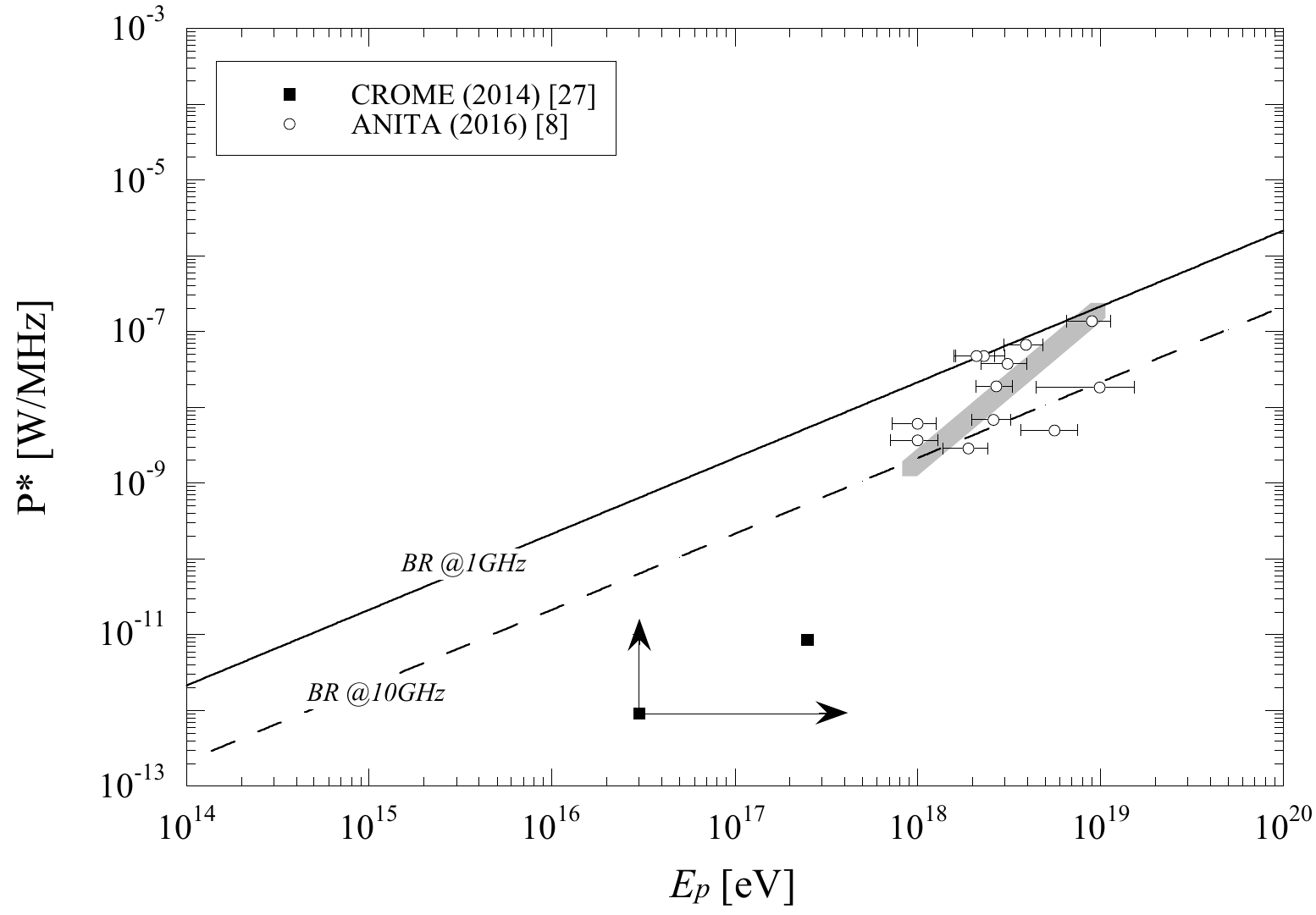}
\caption{Experimental data of the emitted RF power as a function of the cosmic ray energy in the  1--10~GHz range. The grey band is the fit of the ANITA data according to $P^*\propto E_p^2$. Also plotted are the BR emissions, according to eq.~(\ref{eq:BR}), calculated at 1 and 10~GHz. Arrows indicate that the point is a lower limit. 
}
\label{fg:range3}
\end{figure}

For the purpose of this paper, we do not use a Monte Carlo simulation of the EAS, but refer to experimental results. 
We looked for literature data where both the primary cosmic ray energy and the RF emitted power, or energy, or electric field, are simultaneously measured. The primary cosmic ray energy is measured with traditional methods, i.e., scintillators or particle counters, except for the ANITA experiment. Remarkably, few data with such features are available, despite the subject has been investigated for more than half century.  
The oldest ones suffer of poor detector calibration and
 sometimes measures represent only lower limits.

In general, the passage from the locally measured electric field (or power) to the total  power emitted by the EAS introduces uncertainties which can be hardly evaluated in a rigorous way. What is relevant here is to hit the right order of magnitude, with the purpose to justify why the BR radiation has not been detected so far.

We list in Appendix A the literature  data  and how we use them to calculate the RF total emitted power.

Different cases can be encountered. Sometimes the RF electric field $\mathbb{E}^*$ is measured and evaluated at the antenna dipole at a distance $r$  from the shower axis. It has been shown \cite{LOPES2010}  that $\mathbb{E}^*(r) = \mathbb{E}^*(0)\exp(-r/r_0)$, with $r_0$ = 154~m.  
Since the power is: 
$$
 P = \frac{1}{\Delta T}\frac{\epsilon_0}{2}\int(\mathbb{E}^*\Delta\nu)^2dSdz
 $$ 
 where $\epsilon_0$ is the vacuum permittivity and $\Delta T = L_{dish}/c$ for an uniform charge distribution along $z$, the power per unit frequency $P^*$ is 
\begin{equation*}
    \label{eq:pstar}
P^* = \pi\epsilon_0c\Delta\nu \int_0^{\infty}\mathbb{E}^*(r,\nu)^2rdr =\pi\epsilon_0c\Delta\nu\mathbb{E}^*(0)^2~\frac{r_0^2}{4} 
\end{equation*}
\noindent
assuming that the frequency dependence of $\mathbb{E}^*(r,\nu)$  can be neglected within 
$\Delta\nu$.

Some experiments express the detected RF power in term of signal-to-noise ratio with RF noise characterized by the noise temperature $T_{noise}$.  Then the noise power in the bandwidth $\Delta\nu$ is $P_{noise} = K_BT_{noise}\Delta\nu$, where $K_B$ is the Boltzmann's constant.

Other experiments express  the electric field normalized to the antenna area.

In the last two cases, the estimation of  the total power radiated by the EAS requires some care. 
The footprint of the RF radiation on the earth surface is different according to the emission frequency, because of the presence, at high frequency, of the time compression mechanism. 
At frequency below 100 MHz, where the compression  is unimportant, the footprint for a vertical EAS is a circle of radius $R\approx 100$~m,
since the altitude of the maximum shower development $H_{max}$ is about 6~km. At higher frequency, instead, the emission is inside a ring whose radius and radial aperture is defined by the Cherenkov cone, and the footprint area is $S=2\pi H_{max}^2\psi_C\Delta\psi$.

A comment must be done about the ANITA data, collected in their first flight at about 36 km altitude. The RF radiation is detected after it has been reflected by the Antarctic ice sheet and measured in different frequency range from 300 MHz to 1.1 GHz. The energy of the primary cosmic ray is estimated from the same data by modelling the radio emission with a Monte Carlo simulation and not measured by a different detector, as done by all other experiments here considered. 

The results are summarized in figures \ref{fg:range1}, \ref{fg:range2}, and \ref{fg:range3}, where the RF emitted power is plotted as a function of the energy of the primary cosmic-ray in three frequency ranges.

In the 30--100 MHz region (figure~\ref{fg:range1}), the 4 datasets from the most recent experiments (LOPES \cite{Apel2014,LOPES_calibr}, YAKUTSK \cite{YAKUTSK2015}, AERA \cite{AERA2015}, and TUNKA-REX \cite{TUNKAREX_NIM2015}) agree with each other within a factor of 3 and follow the expected quadratic dependence on $E_p$.  Although the power from AERA experiment is calculated in a different way than the other three experiments (see item iv) in the Appendix), the nice compatibility of all final results suggests that the approximations and assumptions done are reasonable. Old data have relevant uncertainties but are all compatible with the extrapolated trend. 

In the 100-500 MHz region (figure~\ref{fg:range2}),  the ANITA data follow the quadratic law. Old data are scattered and barely consistent with the previous ones.

In these two frequency regions, the BR emission calculated with eq.(\ref{eq:BR}) is overwhelmed by the coherent emission in the explored $E_p$ ranges and therefore escaped detection. In the 30--100 MHz range, BR becomes dominant  for $E_p \lesssim 5\cdot10^{16}$~eV. Between 100~MHz and 1~GHz, the cosmic ray energy for equal power emission increases to $E_p \approx 10^{18}$~eV. Therefore,  
we can reasonably infer that the BR emission model described in section~\ref{sec:BR}  is compatible with those experimental data.

In the highest frequency range (1--10 GHz) (figure~\ref{fg:range3}), populated with the ANITA and the CROME data, respectively at 1~GHz and 3.8~GHz (central frequency),  the BR model is apparently incompatible with the experimental results. We point out that the two datasets are not consistent with each other. 
The CROME data seem to scale with $E_p$ in accordance with the ANITA trend, but a further factor $10^4$--$10^5$ on the yield must be expected, because of the dependence (\ref{eq:EvsNu}) of the electric field on the frequency. Note moreover that CROME lacks of a full calibration of the experimental apparatus. These facts cast some
doubts on the absolute scale of the two experiments, and we are not confident to use them to draw conclusions about the BR emission.

\section{A new idea to detect astronomical $\gamma$-rays}
\label{sec:newidea}

As described before, the BR is emitted with a collimated angular shape, similarly to the Cherenkov radiation. Because of this aiming capability, the BR is suitable, in principle, to detect cosmic rays coming from point sources. 
We examine here the possibility to exploit the BR  for the detection of  astronomical $\gamma$-rays with energy above 100~GeV. This novel technique would accompany the UV Cherenkov telescopes which are running nowadays or planned for the future \cite{MAGIC,HESS, VERITAS, CTA}. 
 
 The minimum detectable power (per unit frequency) $\Delta P_{min}^*$  is related to the total noise temperature $T_{noise}$ by the formula \cite{kraus}:
\begin{equation*}
 \Delta P_{min}^* = \frac{2k~K_BT_{noise}}{\sqrt{\Delta\nu~t_{meas}}} 
 \end{equation*}
where $k$ is the sensitivity constant, which depends on the type of receiver, and it is equal to 1 for total power detector, and  $t_{meas}$ is the measurement time. In our case, $t_{meas}$ is the duration of the electromagnetic flash, and approximately $t_{meas} = L_{dish}/c \approx 10$ ns.
$T_{noise} = T_{el} + T_{SKY}$, sum of the receiver electronic noise ($T_{el}$) and of the unavoidable sky background $T_{SKY}$. $T_{SKY}$ is minimum in the range from 1 to 10~GHz, where it counts about 3~K. At lower frequency, $T_{SKY}$ increases rapidly, being around $10^3$~K at 100~MHz and $10^5$--10$^6$~K at 10~MHz \cite{kraus}. Note also that, since the bandwidth $\Delta\nu \propto \nu$,  $\Delta P_{min}^* \propto T_{noise}/\sqrt{\nu}$ . This explains why the highest sensitivity is reached in the microwave (MW) frequency region, i.e., for $\nu \gtrsim 1$~GHz.

To extrapolate, for example, to 100 GeV the experimental data at about 50~MHz (see figure~\ref{fg:range1}), which show the highest intensity among all experimental results, the RF emission must be diminished by a factor $(10^{17}$eV/$10^{11}$eV$)^2 = 10^{12}$, resulting in $P^*\approx10^{-19}$~W/MHz, well below the minimum detectable power at that frequency ($2\cdot10^{-12}$~W/MHz). At $\nu \geq 1$~GHz, where the sensitivity is maximal, and neglecting $T_{el}$, any coherent  signal, extrapolated from the experimental data in figure \ref{fg:range3}, is many decades lower than $\Delta P_{min}^*$. Therefore  coherent emission cannot be exploited for the detection of low-energy EAS in all frequency ranges.

We consider now the BR emission, which  scales linearly with the primary cosmic ray energy. Introducing the parameter $f$, which quantifies the fraction of MW signal collected by the receiving apparatus,
the signal-to-noise ratio is {\it SNR} $= f\cdot P^*/ \Delta P_{min}^*$. Recalling eq.~(\ref{eq:BR}), 
\begin{equation}
\label{eq:SNR}
{\it SNR} = \frac{a}{2kK_B}\frac{f\cdot E_p}{T_{noise}}\frac{\sqrt{\Delta\nu~t_{meas}}}{\nu}
 \end{equation}
{\it SNR} can be optimized by increasing the bandwidth and using the lowest possible frequency. 
In the frequency range 1--10~GHz, where   $\Delta\nu\approx0.1~\nu$,  ${\it SNR}  \propto f/\sqrt{\nu}$. At $\nu$ = 1 GHz 
\begin{equation*}
{\it SNR}  = 0.78~{\rm K\cdot GeV^{-1}}~\frac{f\cdot E_p}{T_{noise}}
 \end{equation*}
According to  model \cite{noi}, the radiation is emitted in the forward direction with a  small aperture angle. 
An antenna dish like the mirror of the UV Cherenkov telescopes discussed below can collect about $1\%$  of the emitted power.
With $T_{noise}=10$~K, the energy threshold for a detection with {\it SNR}~=~1 is  $E_p \approx 1$~TeV.
Measurements could also be performed with astronomy radiotelescopes, which have larger dishes, up to 100~m diameter,  for which $f \approx 0.1$ and the energy threshold lowers to a hundred of GeV.

On the opposite, in the 10--500~MHz range the sky noise is too high, making hopeless the detection of  $\gamma$-ray EAS. 
 
In principle, the best possible (and perhaps definitive) verification of this idea is to measure the MW emission from $\gamma$-ray EAS in coincidence with an alternative and trustable measure of the same shower. 
This can be done by aiming at a $\gamma$-ray source (for example, the Crab Nebula) with the same telescope for the MW and the UV radiation, and detecting simultaneously the two signals.

\begin{figure}[!t]
\centering
\includegraphics[scale=0.4]{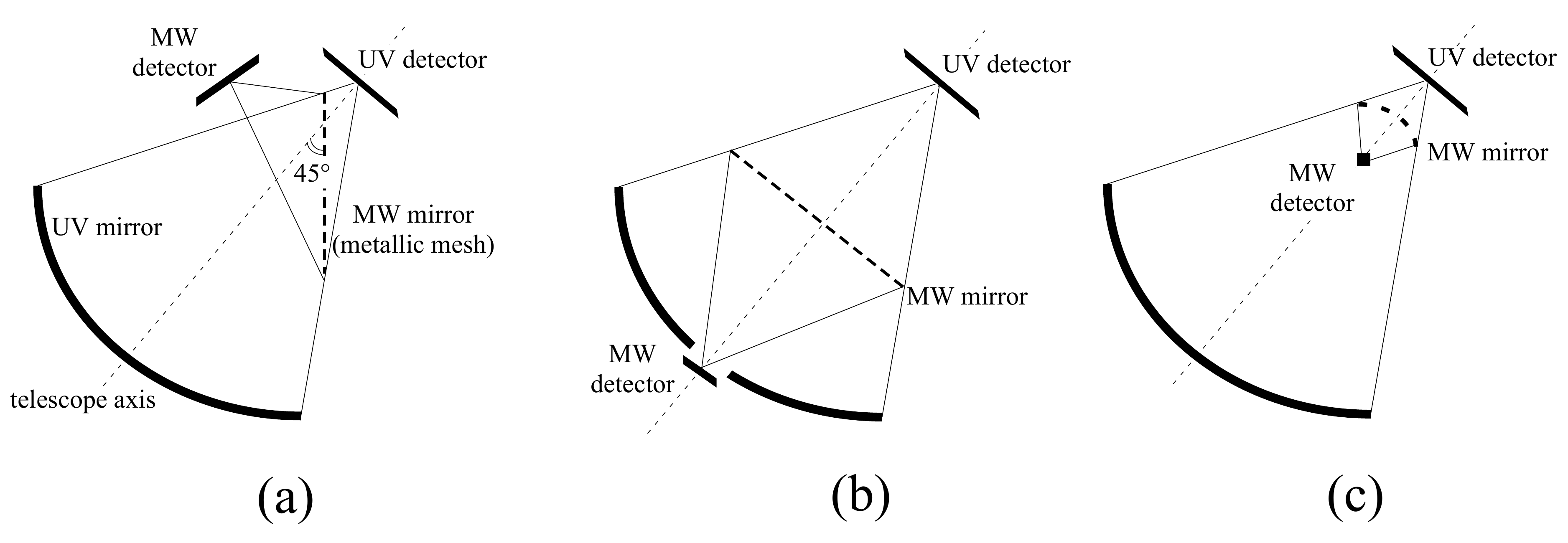}
\caption{
Three possible experimental configurations to detect simultaneously UV and MW signals from the same $\gamma$-ray EAS using an UV Cherenkov telescope, as discussed in the text (not to scale).
}.
\label{fg:exp}
\end{figure}

Suppose a telescope with an aluminium mirror capable to reflect both UV and MW radiation (Al thickness $\sim 1$~mm), with diameter of the order of 15~m. MW can be reflected also by mean of a planar or curved metallic  grid, with a mesh dimension of the order of a few mm, and a wire diameter of the order of few hundreds $\mu m$. Such grid is transparent for the UV photons (the transparency is simply the geometrical open area, $> 90\%$), therefore it can be placed inside the Cherenkov telescope, intercepting the optical path, without disturbing the UV measurement, while the MW reflectivity is about $100\%$. Some possible configurations are sketched in figure~\ref{fg:exp}:
\begin{enumerate}
\item (figure~\ref{fg:exp}a) a MW planar mirror is inclined at $45^\circ$ with respect the telescope
axis, to reflect the MW radiation away from the telescope optical path. The MW detector is off-axis.
\item (figure~\ref{fg:exp}b) the MW detector is placed at the centre on the mirror, if a hole is present. A planar MW mirror is inserted to reflect MW towards the detector. In case of f/1 mirror, the MW mesh  is at half focal length and has diameter equal to $D/2$, obscuring 1/4 of the incident MW radiation. 
\item (figure~\ref{fg:exp}c) a curved (convergent) mirror is placed on the telescope axis to reduce the focal length for the MW radiation. The magnification factor is less than 1 and the MW detector dimension can be reduced, in order to shadow a minimal fraction of the UV detector.
\end{enumerate}

\section{Conclusions}
\label{sec:conclusions}

The detection of astronomical $\gamma$-rays with energy above 100 GeV seems possible by exploiting the BR emission at 1--10 GHz, as long as the emission  from the EAS follows the model here depicted. Such model is based on experimental data at very low energy and requires an experimental verification at high energy. A possible setup for such a test has been described.  The test is well worth doing because of the potential significant impact on the  $\gamma$-ray astronomy technique:
\begin{enumerate}[i)]
\item since the MW emission and detection are almost insensitive to light and weather conditions, the observation duty cycle, limited nowadays to less than 1000~hours/year for the UV Cherenkov telescopes, would be dramatically boosted;
\item the requirements for the location of the observatory are loose;
\item the MW technique requires a cheaper hardware than that used by actual UV Cherenkov observatories.
\end{enumerate}


\newpage
\appendix
\section{Appendix: literature data}
\label{sec:appendix}

Here we report the experimental data which enter in the figures~\ref{fg:range1}, \ref{fg:range2}, \ref{fg:range3} and how we calculated the total emitted power.

The most recent data are:

\begin{enumerate}[i)]

\item LOPES data in the range 43 to 76 MHz \cite{Apel2014}. We use their fit from the experimental data in the range 0.6--40~$\mu$V/m/MHz: $ (E_p/10^{17}~{\rm eV}) =(0.134 \pm 0.005)\cdot \mathbb{E}^*(d_0)~ \mu$V/m/MHz, where the electric field is calculated at the distance $d_0=70$~m from the shower axis.
Later on, the result has been corrected because of a better absolute calibration of the antenna array \cite{LOPES_calibr}. The effect is that $\mathbb{E}^*$ has to be decreased by a factor $2.6 \pm 0.2$, which has been considered in our final calculation.

\item CROME data in the range 3.4 to 4.2 GHz \cite{Smida2014}. The power lower limit is defined by the trigger threshold, which has signal-to-noise ratio $\ge 8$~dB, for a noise temperature $T_{noise}=90$~K. Such threshold selects cosmic rays with energy above $3\cdot10^{16}$~eV. The primary with the highest energy signal ($E_p=2.5\cdot10^{17}$~eV) has emission with signal-to-noise ratio = 17.7~dB, detected at a distance $d_0=120$~m from the shower axis, and inclination $\theta=5.6^{\circ}$. Due to the high frequency, we suppose that the emission occurs within a ring of area $S=2\pi H_{max}^2 \psi_C\Delta\psi/\cos\theta$, with $\psi_C=1^{\circ}$, $\Delta\psi=0.1^{\circ}$.

\item YAKUTSK data at 32~MHz, bandwidth 8~MHz \cite{YAKUTSK2015}. We use their fit from the experimental data in the $E_p$ range from  $6\cdot10^{16}$~eV  to $4\cdot10^{18}$~eV for vertical showers at zero distance from the shower axis: $\mathbb{E}^*(0)=(15\pm1)\cdot (E_p/10^{17}~{\rm eV})^{0.99\pm0.04}~\mu$V/m/MHz.

\item AERA data  in the range 30 to 80 MHz \cite{AERA2015}. The total radiated energy is estimated as: $(15.8\pm0.7 {\rm (stat)} \pm 6.7 {\rm (sys))~MeV} \cdot (E_p/10^{18}~{\rm eV})^2$, when shower axis is parallel to the earth magnetic field, which has local intensity of 0.24~G. We suppose an emission time $\Delta T = 10$~ns. 

\item TUNKA-REX data in the range 30 to 76 MHz \cite{TUNKAREX_NIM2015}. The trend is extrapolated from the data shown in the figure~12. The electric field intensity was measured at the distance $d_0=120$~m from the EAS axis.

\item ANITA data at 300 MHz \cite{ANITA2015}. The  amplitude $A$ of the electric field is reported in units $\sqrt{{\rm pW/MHz/m}}$. Schematically, the radiation is emitted at an altitude about $H_{max}$, and detected by the apparatus at 36~km elevation, after it has been reflected by the Antarctic ice sheet. The Cherenkov emission angle $\psi_C$ and angular width $\Delta\psi$ with respect to the direction of the primary, inclined at an angle $\theta$, are also measured. The RF radiation is emitted within a ring of area $S = 2\pi H^2 \psi_C \Delta\psi/\cos\theta$, where $H \approx 42$~km. The reflectivity $\rho$ of the ice has also to be taken into account. Therefore $P^* = A^2\cdot S/\rho$.

\item ANITA data at 1 GHz \cite{ANITA2015}. The  amplitude $A($300MHz)  of the electric field at 300~ MHz is reported. The amplitude $A(\nu)$ scales with $\nu$ as eq.~(\ref{eq:EvsNu}), with $\nu_{\tau}$ given event by event. Once $A$(1GHz) is obtained, the emitted power is calculated as the previous entry.

\end{enumerate}

\noindent
Going to less recent measurements:

\begin{enumerate}[i)]
\setcounter{enumi}{7}

\item Jelley et al. data at 44 MHz \cite{Jelley1965}. The total energy collected by the antenna is reported. The total power is then computed considering an antenna area of 1800~m$^2$, an emission time $\Delta t = 10$~ns, and that the RF footprint surface is a circle of radius $R =100$~m.

\item Jelley et al. data at 44 MHz \cite{Jelley1966}. As previous entry, except that the antenna surface was 1~m$^2$.

\item Vernov et al. data at 30 MHz \cite{Vernov1968}. The number $\mathcal{N}$ of secondaries on the earth surface was measured, and we estimated the cosmic ray energy using the relationship $\mathcal{N} = 0.66\cdot E_p{\rm (GeV)}^{1.03}$ \cite{Matthews}. The RF power density was reported at a distance $d_0=150$~m. Since the power $P\propto \mathbb{E}^2$ and $\mathbb{E}(r)=\mathbb{E}(0)\exp(-r/r_0)$, then the total power is obtained as $P= \int_0^{\infty}P(r)rdr=P(0)\int_0^{\infty}r\exp(-2r/r_0)dr$.

\item Charman data at 44 MHz \cite{Charman1968}. With $T_{noise}=5\cdot10^3$~K, signals had a signal-to-noise ratio > 4. The RF footprint has an area $S=\pi R^2/\cos\theta$, with $R=100$~m and $\theta=70^{\circ}$ (EAS axis inclination).

\item Charman and Jelley data at 200 MHz \cite{Charman1969}. The energy received by the antenna ( 1~m$^2$ area) is reported, for $E_p=3\cdot10^{17}$~eV. The total power is estimated assuming that the RF footprint surface is a circle of radius $R=$100~m, and an emission time $\Delta T=10$~ns.

\item Charman data at 500 MHz \cite{Charman1970}. The RF energy per unit surface is reported. The total RF energy is calculated assuming that the RF footprint surface is a ring with surface $S=2\pi H_{max}^2\psi_C\Delta\psi$ ($\psi=1^{\circ},\Delta\psi=0.1^{\circ}$), and an emission time $\Delta T=10$~ns.

\item Mandolesi et al. data at 46, 65, and 110 MHz  \cite{Mandolesi1976}. The number $\mathcal{N}$ of secondaries on the earth surface was measured, and we estimated the cosmic ray energy using the relationship $\mathcal{N} = 0.66\cdot E_p{\rm (GeV)}^{1.03}$.

\end{enumerate}


\end{document}